\documentclass[a4paper,11pt]{article}

\usepackage{jcappub} 

\title{\boldmath Scattering of the UHECR at small pitch angle by damped plasma waves}

\author[1]{M. Vukcevic\note{Corresponding author.}}

\affiliation[]{Astronomical Observatory Belgrade,\\Volgina 7, Serbia}

\emailAdd{vuk.mira@gmail.com}

\abstract{In spite a lot of theoretical and experimental effort that has been achieved in ultra-high energy cosmic ray (UHECR) scattering research in last few decades, some questions remain unanswered, or partially answered. Two of them, that will be in the focus of this paper are: possible source of UHECRs and the acceleration mechanism of cosmic rays beyond PeV energies. Small pitch-angle scattering of UHECRs and possible confinement has been investigated using quasilinear theory in order to analytically calculate pitch-angle Fokker-Planck coefficient. CR particles resonantly interact with oblique low frequency damped waves.  We show that the resonance function is broadened due to damping effects and this result is compared with the nonlinear broadening. Unlike the case of purely parallel (or antiparallel) propagating waves in slab turbulence, the presence of the compressive magnetic field component of oblique fast-mode waves allows the cosmic ray particles to resonantly interact with these waves through the n = 0 resonance, together with gyroresonance, which strongly influence the Hillas limit. The derived results can be used to compute the parallel mean free path for all forms of the turbulence spectrum; it has been applied on the transport and propagation of CRs close to ultra-high energies in the Galaxy.
An accurate understanding of particle acceleration in astrophysical sources could help to interpret eventual transition from Galactic to extragalactic origin of cosmic rays, if any,  and the shape of the UHECR spectrum at the highest energies.}

\begin{document}
\maketitle
\flushbottom

\section{Introduction}
\label{sec:intro}

An open question regarding the origin of UHECRs has not been answered yet. The most intriguing are candidates within our host galaxy that can accelerate the protons, providing energies up to $10^{19}-10^{20}$eV. We focus on the possible sites inside the Milky Way because of the well known Greisin, Zatsepin and Kuzmin (GZK) cutoff restriction \citep{Greis66}. This particular restriction states that protons of these energies would interact with microwave background radiation through photo-pion reactions and lose energy on a length scale of about few Mpc which is relatively short distance in cosmological terms (radius of a flat disk of our Galaxy is approximately 15 kpc). On the other side, there is a requirement on the maximum size of the acceleration site, known as a Hillas limit 
\citep{Hill84}. It states that particles can be confined within the accelerator domain as long as their Larmor radius $R_{L}$ is smaller than the size of the accelerator. Scattering of the particles is effected by the field perturbations of wavelength proportional to $2\pi R_{L}$, or equivalently of wavenumber $k_{min}$, where $k_{min}R_{L}$ defines particle energy limit. Apart of the size of accelerator,
in the Hillas limit exists the value of the magnetic field due to $R_{L}$. 
The magnitude and structure of the magnetic field are a major uncertainties in the model (the uncertainty of the regular structure, of the random magnitude and structure). Huge improvement has been made during past decades \citep{Kuz21} (section 5.2), giving a possibility to improve Fermi acceleration contribution via turbulent medium considering more generalized case than Hillas did, namely examining the oblique turbulence model instead of slab one. Resonant wave number particularly depends on the turbulence model and in the case of small pitch-angle contribution is undefined for slab turbulence model. Discard from the slab to oblique model allows scattering even in small pitch-angle enhancing particle energy up to ultra-high values. In certain turbulent conditions one can expect that it is able to provide particle energy up to ultra high values since the mean free path of the particle can achieve even $25R_{L}$ \citep{Hill84}. In this paper we will reinvestigate these certain conditions involving damping effects that can enhance energy filed which contains resonant wavelengths, or resonant wave numbers, responsible at very small pitch-angle scattering process. That is why we think that the well known Figure 1 of Hillas \citep{Hill84} should be reexamined by employing oblique plasma wave turbulence model in the case of interstellar medium. Implication of the resonant wavenumber on the particle energy limit will be discussed in details in Section 2.1.
 
To reach a more complete view of the UHECR dynamics, cross-correlation studies have been performed within the Auger and TA collaborations, as well as by independent groups. Models often assume that the UHECR source distribution follows the distribution of luminous matter in the nearby Universe, based on radio, infrared, X-ray and gamma ray observations. These models account for the expected energy losses and deflections of UHECRs during their extragalactic propagation \citep{Kas08, Oik13, Aab05b}. 
The arrival directions of UHECRs favor starburst galaxies, AGN and radio galaxies as likely sources.
Powerful FRII galaxies are rare within the local GZK volume, while FRI radio galaxies are more populous \citep{Mass09}. Further evidence pointing toward starburst galaxies and away from radio galaxies is the analysis of \citet{Abasi20} which does not find a a statistically significant association of TA arrival directions with starburst galaxies. However, even though both collaborations report statistical significance of 5 sigma, it cannot be understood as a proof for extragalactic origin because of two reasons: first, as the number of detections increases, the anisotropy amplitude decreases \citep {Abreu10}, and second, reported correlation significance can be taken only as a necessary but not sufficient condition.

Since results of these collaborations have to be taken within certain statistical limitations, we keep searching candidates within our own galaxy by reinvestigation of the possible acceleration mechanism.

The propagation and acceleration of charged cosmic rays are examples of the application of diffusion theory, relevant for different physical systems: shock waves 
\citep{Zank06}, the interstellar medium 
\citep{Sch02}, the heliosphere 
\citep{Alania08} and the solar corona 
\citep{Gki07}.
The parallel mean free path $\lambda$ of the charged particle is related to the parallel spatial diffusion coefficient $\kappa$ as $\lambda= 3\kappa/v$ and can be expressed by the integral over the inverse of the pitch-angle Fokker-Planck coefficient $D_{\mu\mu}$ 

\begin{equation}
\label{ a10}
\kappa_{z z}=v\lambda/3=\frac{v^{2}}{8}\int_{-1}^{1}d\mu \frac{(1-\mu^{2})^{2}}{D_{\mu \mu}(\mu)},
\end{equation}

with the pitch-angle cosine $\mu= v_{||}/v$ and the particle velocity $v$. 
The application of perturbation theory also known as quasilinear theory (QLT \citep{Jok66}) was the first attempt to derive parameter $D_{\mu\mu}$. 
QLT was not able to describe pitch-angle scattering at $90^{0}$ ($\mu = 0$) correctly. 

This problem was investigated in numerous papers (see, e.g., \citep{Owens74, Gold76, Jon78}), where QLT has been improved by replacing unperturbed orbits by more appropriate models. A second order quasilinear theory (SOQLT, \citep{Sha05}) was developed providing moderate agreement with test-particle simulations. 
However, it has been noticed that resonance function broadening can be achieved modeling MHD turbulence by relevant plasma wave modes  considering damping effects. Relevant modes assure non-zero $D_{\mu\mu}$ parameter at $\mu=0$ and modification of the Hillas limit. 

Therefore, according to the current understanding the relativistic charged particles in space sites within our own galaxy are confined and accelerated by resonant interactions in the interstellar medium plasmas. It has been already shown by \citet{Vukcevic07} that resonant interactions of CR with relevant plasma modes are able to confine and accelerate protons up to four orders of magnitude higher energies then defined by Hillas limit. In this paper, we reinvestigate Hillas limit expression, altogether with damping effects of the plasma wave turbulence. We investigate a fundamental problem of CR diffusion theory, scattering at small pitch-angle, employing oblique propagation of damped magnetosonic waves, where slab turbulence model leads to an infinitely large mean free path. We compare relevance of resonance function broadening due to damping  with those induced by nonlinear effects. Plasma wave turbulence is more realistic than magnetostatic turbulence used in nonlinear approach. Even more, plasma wave approach is self-consistent which means that conditions $V_{A}/v$ and $\delta B/B_{0}$ cannot be discussed independently. They are dependent via Lorentz force that is used for calculation of relevant component of the fluctuating force term, namely $\dot{\mu}$. Departure from pure parallel or antiparallel propagation of the waves with respect to ambient magnetic field will allow modification of the Hillas limit enhancing it up to ultrahigh energies of particles.
As far as the anisotropy is concerned, it is directly proportional to the mean free path \citep{Sch89}

\begin{equation}
\label{ a10}
\delta=\frac{\lambda}{3[z]},
\end{equation}
where $[z]$ is characteristic spatial gradient of CR density and according to \citet{Strong96} its value is $\simeq 2kpc$.

In this paper we reinvestigate
the Hillas limit; how it is affected if we discard
the assumption of purely slab plasma waves, and how it will be influenced by damping effects inclusion.
The first assumption on the slab turbulence will result in the changed resonant wavenumber, wile the second one, regarding damping effects, will allow transit-time damping even at small pitch-angle scattering.
There is observational evidence that obliquely propagating
magnetohydrodynamic plasma wave exists in the interstellar medium 
\citep{Lit01, Cho02}.
It has been pointed out by
\citep{Sch98} that oblique propagation
angles of fast magnetosonic waves leads to an order of magnitude quicker
stochastic acceleration rate as compared to the slab case, since the
compressional component of the obliquely propagating fast
mode waves allows the effect of transit-time damping acceleration of particles. 
Here we show that the obliqueness of
fast mode wave propagation and damping effects via broadening resonance function
modify the resulting parallel spatial diffusion coefficient
and the limit. Moreover we will show that the maximum wavelength
$L_{\rm max}$ of isotropic waves does not have such a strong effect on the
maximum particle rigidity as in the slab case.

\section{Method} \label{sec:meth}

The anisotropy and confinement of CR is defined by the parallel spatial diffusion coefficient or equivalently by mean free path and corresponding Fokker-Planck coefficient $D_{\mu\mu}$, as it is given by Eq. (1.1). The easiest method to derive the coefficient $D_{\mu\mu}$ is to apply quasilinear perturbation theory. In that theory particle velocities and trajectories are replaced by the unperturbed orbits 
\citep{Jok66}. Appart of QLT, further assumption is related with the turbulence model. In this work we consider turbulence as a mixture of relevant plasma wave  modes which asks for time integration of the Fokker-Planck coefficient, on contrary to the magnetostatic turbulence \citep{Teufel03}. 
As far as the geometry of the turbulence is concerned, it has been widely used slab model \citep{Hill84}, applying the correlation tensor that is just special case of the isotropic one (\citep{Vukcevic07}, Eq. (23)). It means that we consider waves that can propagate at some angle $\theta$ with respect to ambient magnetic filed. Implication of the oblique plasma wave propagation on the Hillas limit with no damping effects (resonance function is rather sharp, represented by delta-function) has been discussed by \citet{Vukcevic07}. 

Since the largest value of the mean free path is defined by the smallest value of the $D_{\mu\mu}$ coefficient, we deal with the case $D_{\mu\mu}(\mu \rightarrow 0)$:

\begin{equation}
\label{ a10}
\lambda=\frac3{v}{8}\int_{-1}^{1}d\mu \frac{(1-\mu^{2})^{2}}{(D^{G}_{\mu \mu}(\mu \rightarrow 0)+D^{T}_{\mu \mu}(\mu \rightarrow 0))},
\end{equation}

where supscript $G$ denotes gyroresonance contribution and $T$ denotes transit-time contribution.

This scenario in the case of undamped waves has been investigated by \citet{Vukcevic07} and it was derived that the only contribution for small $\mu$ comes from gyreroresonant interaction $(n\neq0)$. In such a case particles resonantly interact with particular single wavenumber. However, inclusion of damping would broaden resonance function allowing resonant interaction with all wave-numbers, so that transit-time contribution to Fokker-Planck coefficient will be nonzero. This case will be discussed in Sec. 2.2.

\subsection{Dispersion relation and resonant wave number}

In order to calculate $D_{\mu\mu}$ coefficient, we follow the approach for the electromagnetic turbulence that
represents the Fourier transforms of the magnetic and electric field fluctuations as superposition of
$N$ individual weakly damped plasma modes of frequencies

\begin{equation}
\label{ }
\omega=\omega_{j}(\textbf{k})=\omega_{R,j}(\textbf{k})-i\gamma_{j}(\textbf{k}),
\end{equation}
$j=1,...N$, which can have both the real and imaginary parts with $|\gamma _j|<<|\omega _{R,j}|$. Damping of the waves is counted with the a positive $\gamma^{j}>0$.

For obliquely propagating MHD waves relevant are the fast mode and shear Alfven wave. It has been already shown that the gyroresonant contribution to the Fokker-Planck coefficient from Alfven waves can be neglected since the main contribution comes from fast mode waves (\citep{Vukcevic07}, Eq. (48)). Here, we underline the implication of the Hillas assumption of the resonant wave number on the energy limit.

Dispersion relation for fast magnetosonic waves read as:

\begin{equation}
\label{ }
\omega _R\simeq ikV_A
\end{equation}

describing forward ($i=1$) and backward ($i=-1$) moving fast mode waves.
Resonant wavenumber in the gyroresonance case for an isotropic turbulence reads as:
\begin{equation}
\label{ }
k_{i}^{g}(\mu)=\frac{n\Omega}{v(\mu\eta-i\epsilon)}=\frac{n}{R_{L}(\mu\eta-i\epsilon)}
\end{equation}

where $R_{L}=v/\Omega$ is gyroradius. Resonant wavenumeber in the slab turbulence is 

\begin{equation}
\label{ }
k_{i}^{g}(\mu)=\frac{n}{R_{L}\mu}
\end{equation}

and it is often approximated as

\begin{equation}
\label{ }
k_{i}^{g}(\mu)\simeq\frac{n}{R_{L}},
\end{equation}

which actually is undefined at $\mu=0$ (see Eq. (2.5)). 

At $\mu=0$ resonant wavenumber for isotropic turbulence (2.4) will be as follows:
\begin{equation}
\label{ }
k_{i}^{g}(\mu=0)=\frac{n}{R_{L}\epsilon},
\end{equation}

that has no singularity at $\mu=0$, it is independent of the magnetic field strength ($k_{i}^{g}(\mu=0)=\omega_{p,i}/\gamma c=1.88\times10^{-6}\sqrt{n_{e}}/\gamma$) and it is by factor of $\epsilon^{-1}=c/V_{A}$ larger than the Hillas limit \citep{Hill84}.

\subsection{Implication of damping}

In this section we investigate how inclusion of damping would influence Fokker-Planck coefficient, and consequently the mean free path of UHECR. 
The non-vanishing parallel magnetic field component $B_{\parallel }\ne 0$ of fast mode waves
allows transit-time damping interactions with $n=0$. It has been pointed out by \citet{Sch98}
that this transit-time damping (TTD) contribution provides the
overwhelming contribution to
particle scattering because in this interaction the cosmic ray particle interacts with the whole wave spectrum,
in contrast to gyroresonances that singles out individual resonant wave numbers (see also the discussion in
\citet{Sch02}). 

Damping waves would influence magnetic turbulence tensor as follows:

\begin{equation}
\label{k10}
P_{\alpha\beta}(\textbf{k},t)=\sum_{j=1}^{N}P_{\alpha\beta}^{j}(\textbf{k})
e^{-i\omega_{R,j}(\textbf{k})t-\gamma_j(\textbf{k})t},
\end{equation}

where $t$ is time. The time integration yields following resonance function:
\begin{equation}
\label{k12}
{\cal R}_j(\gamma _j)=\int_0^\infty dt\; e^{-i(k_{\parallel}v_{\parallel}+\omega _{R,j}+n\Omega)t-\gamma _jt}
={\gamma _j(\textbf{k})\over
\gamma ^2_j(\textbf{k})+[k_{\parallel}v_{\parallel}+\omega _{R,j}(\textbf{k})+n\Omega)]^2}.
\end{equation}

In the case of negligible damping $\gamma \rightarrow 0$, reduces the resonance function (\ref{k12}) to sharp $\delta $-functions $\textbf{R}^j(\gamma=0)=\pi \delta(k_{\parallel}v_{\parallel}+\omega_{R,j}+n\Omega))$.

The inclusion of resonance broadening due to wave damping in the resonance function (2.9)
 guarantees that this dominance also holds for cosmic ray particles at small pitch angle cosines
$\mu \le |V_a/v|$, unlike the case of negligible wave damping discussed by \citet{Sch98}. Therefore, in the following we will only take into account the TTD-contribution to particle
scattering and assume $n=0$ both in the resonance function (2.9) and in the calculation of the Fokker-Planck
coefficients. This justified approximation greatly simplifies the evaluation of the Fokker-Planck coefficients.
Since we consider only TTD-contribution, only fast and slow magnetosonic waves are subjects of it. It has been already emphasized that in the case on negligible damping, there is no TTD for shear Alfven waves \citep{Teufel03} and the gyroresonant interactions provided by shear Alfven waves is small compared to the same contribution provided by fast magnetosonic waves \citep{Vukcevic07}. As a consequence, we consider only fast magnetosonic waves in this section, as an example.


The damping of fast mode waves is caused both by collisionless Landau damping and collisional
viscous damping, Joule damping and ion-neutral friction. According to \citet{Spa05} the dominating contribution is provided by  viscous damping. Following the analysis and steps performed by \citet{Vuk13} resonance function reads as:

\begin{equation}
{\cal R}^{jT}_F(\mu)=
{2.9\cdot 10^5\beta V_A^2\sin ^2\theta \over
(2.9\cdot 10^5\beta V_A^2k\sin ^2\theta )^2+[v\mu \cos \theta +jV_A]^2},
\label{d2}
\end{equation}

where $T$ denotes assumption n=0 in Eq. (2.9), while for $\mu=0$ resonance function is

\begin{equation}
{\cal R}^{jT}_F(0)=
\frac{\alpha(1-\eta^{2})}{(\alpha(1-\eta^{2})k)^{2}+V^{2}_{A}}
,
\label{d2}
\end{equation}
with $\alpha=2.9\cdot 10^5\beta V_A^2$, $\beta=c^{2}_{s}/V^{2}_{A}$ and $cos\theta=\eta$.

\begin{figure}
\centering
\includegraphics[width=0.9\columnwidth]{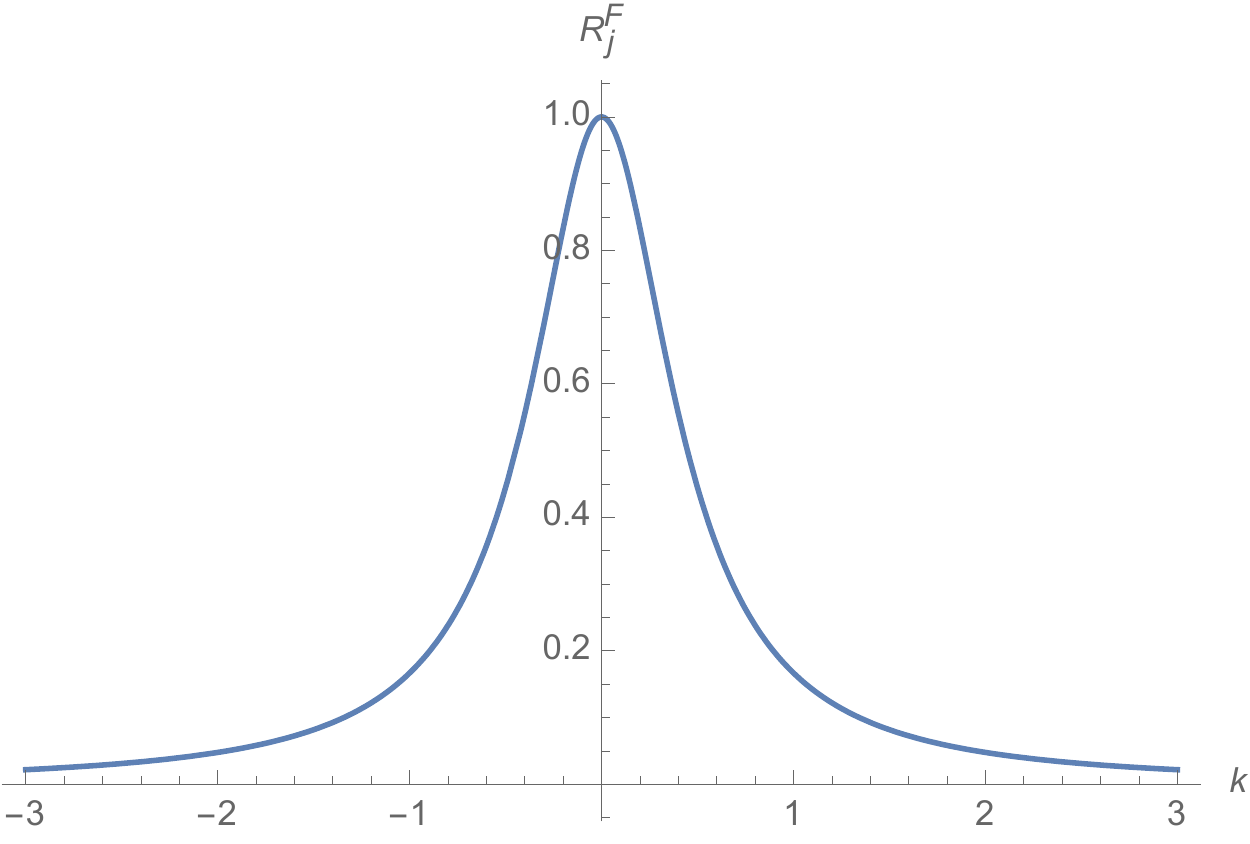}
\caption{The resonance function given by equation (2.11) for typical values in the ISM.}
\label{fig1}
\end{figure}

The resonance function given by equation (2.11) is presented in Fig. 1 for typical ISM values for $\beta$ and $V_{A}$. Comparing our result with the result obtained using second order QLT \citep{Sha09} it can be seen that width of the resonance function for damped plasma wave turbulence is independent of wave number $k$. Also, in the case of plasma wave turbulence, which is more realistic than magnetostatic one, it is not correct to simply separate  velocity and magnetic field contribution since they are not independent. In the plasma wave turbulence the problem is solved self-consistently and time integration is not independent of turbulence geometry; once isotropic tensor (2.8) is involved it is necessary to consider relevant plasma modes together with the dispersion relation (2.3). 

\section{Pitch-angle Fokker-Planck coefficient}

We have already discussed in Sec. 2.2 that inclusion of resonance broadening due to wave damping in the resonance function guarantees dominance of transit-time damping. The main contribution of waves damping comes exactly in the region $|\mu|<\epsilon$ that is relevant in deriving the spatial diffusion coefficient and related mean free path which  are given by the average over $\mu$ of the inverse of $D_{\mu \mu}$. Therefore we can further consider only the case $D_{\mu \mu }(\mu=0)$, which simplifies the analysis enormously. 
Following the procedure of the Fokker-Planck coefficient calculation for damped waves given in \citet{Vuk13}, we derive pitch-angle Fokker-Planck coefficient as follows:

\begin{equation}
D_{\mu \mu }^{F}(\mu=0)\simeq \frac{(q-1)}{\alpha}(k_{min}R_{L})^{q-1}(\frac{\delta B}{B_{0}})^{2}
\int_{k_{min}R_{L}}^{\infty}ds s^{-q}\\
\int _{0}^{1}\, d\eta \; (1-\eta^{4})
J_{1}^{2}(s\sqrt{1-\eta^{2}})\frac{1}{(1-\eta^{2})^{2}s^{2}+\frac{V_{A}^{2}R_{L}^{2}}{\alpha^{2}}},
\label{d14}
\end{equation}

where $s=kR_{L}$. Note that relevant is only the case $\mu<\epsilon$ and since $\mu\rightarrow 0$ integration with respect to $\eta$ goes from 0 up to 1. Also, in the case of UHECR energy limit will be different from the case of positrons treated in \citep{Vuk13} due to different mass of protons end positrons.

\section{Cosmic Ray Mean Free Path for damped FMS Waves}

According to Eq. (1.1) it is possible to calculate mean free path as follows

\begin{equation}
\lambda^{0F}=\frac{3 \kappa}{v}=\frac{3v}{4}\frac{1}{D_{\mu \mu }(\mu=0)}\int _{0}^{\epsilon}\, d\mu \; =\frac{3}{4}\frac{V_{A}}{D_{\mu \mu }^{F}(\mu=0)}=\\
\frac{3V_{A}}{4}\frac{\alpha}{(q-1)}(k_{min}R_{L})^{1-q}(\frac{B_{0}}{\delta B})^{2}\frac{1}{D},
\label{d14}
\end{equation}\\
where it is reasonable to take the boundaries in $\mu$ integration from 0 to $\epsilon$ instead of 0 to 1 since $\mu<<\epsilon$, and 

\begin{equation}
D=\int_{k_{min}R_{L}}^{\infty}ds s^{-q}\int _{0}^{1}\, d\eta \; (1-\eta^{4})
J_{1}^{2}(s\sqrt{1-\eta^{2}})\frac{1}{(1-\eta^{2})^{2}s^{2}+\frac{V_{A}^{2}R_{L}^{2}}{\alpha^{2}}}.
\label{g1}
\end{equation}

Now, we consider two limits: $k_{min}R_{L}<<1$, and $k_{min}R_{L}>>1$, where $k_{min}R_{L}=E$ and is normalized with respect to $E_{c}$, where

\begin{equation}
\label{ }
E_{c}=\frac{k_{c}}{k_{min}}Ac^{2}=2\times10^{5}An^{1/2}_{e}\frac{L_{max}}{(10-100)pc}eV,
\end{equation}
  where $A$ is CR mass, $k_{c}=\Omega_{0,p}/V_{A}=\omega_{p,i}/c$ and $k_{min}=2\pi/L_{max}$.
Evaluation of $D$ in these two limits can be found in \citet{Vuk13}, Appendix A, in details. Note that $k_{c}$ is different for UHECR from CR treated in the cited paper due to different particle mass.\\

$k_{min}R_{L}>>1$:

In this limit we derive

\begin{equation}
D(E>>1)=\frac{2}{5}\frac{10^{-14}}{q}E^{-(q+2)},
\label{d14}
\end{equation}
so that mean free path reads as:

\begin{equation}
\lambda^{0F}(E>>1)=\frac{15 \alpha}{V_{A}}\frac{q}{q-1}(\frac{B_{0}}{\delta B})^{2}10^{14}E^{3}.
\label{d14}
\end{equation}

At relativistic rigidities we find that $\lambda^{0}\sim E^{3}$.\\

$k_{min}R_{L}<<1$:

In this case we derive 

\begin{equation}
D(E<<1)=\frac{1}{3}\frac{1}{q-1}E^{1-q},
\label{d14}
\end{equation}

and consequently 

\begin{equation}
\lambda^{F0}(E<<1)= \frac{9\alpha}{V_{A}} (\frac{B_{0}}{\delta B})^{2}.
\label{d14}
\end{equation}

In this energy limit the mean free path is constant with respect to $E$.
The mean free path is normalized by factor $\lambda_{1}=\frac{9\alpha}{V_{A}} (\frac{B_{0}}{\delta B})^{2}$ which for typical interstellar plasma ($V_{A}=30km/s$, $B_{0}/\delta B\sim 10$ and $q=5/3$) is $2\times10^{15}cm$. In Fig. 2 is presented mean free path for damped and undamped case. It is very important to notice that energy limit is not changed compared to undamped case but the value of the mean free path is two orders of magnitude smaller then in the undamped case, which makes damping plasma turbulence very efficient mechanism in confinement of the UHECR. As we have already mentioned plasma wave turbulence is more effective for small pitch-angle scattering comparing to nonlinear theory (see forth case in Table 2 of \citep{Sha09}; nonlinear plasma wave regime is exactly damping effects relevant for small $\mu$ due to broaden resonant function).

\begin{figure}
\centering
\includegraphics[width=0.9\columnwidth]{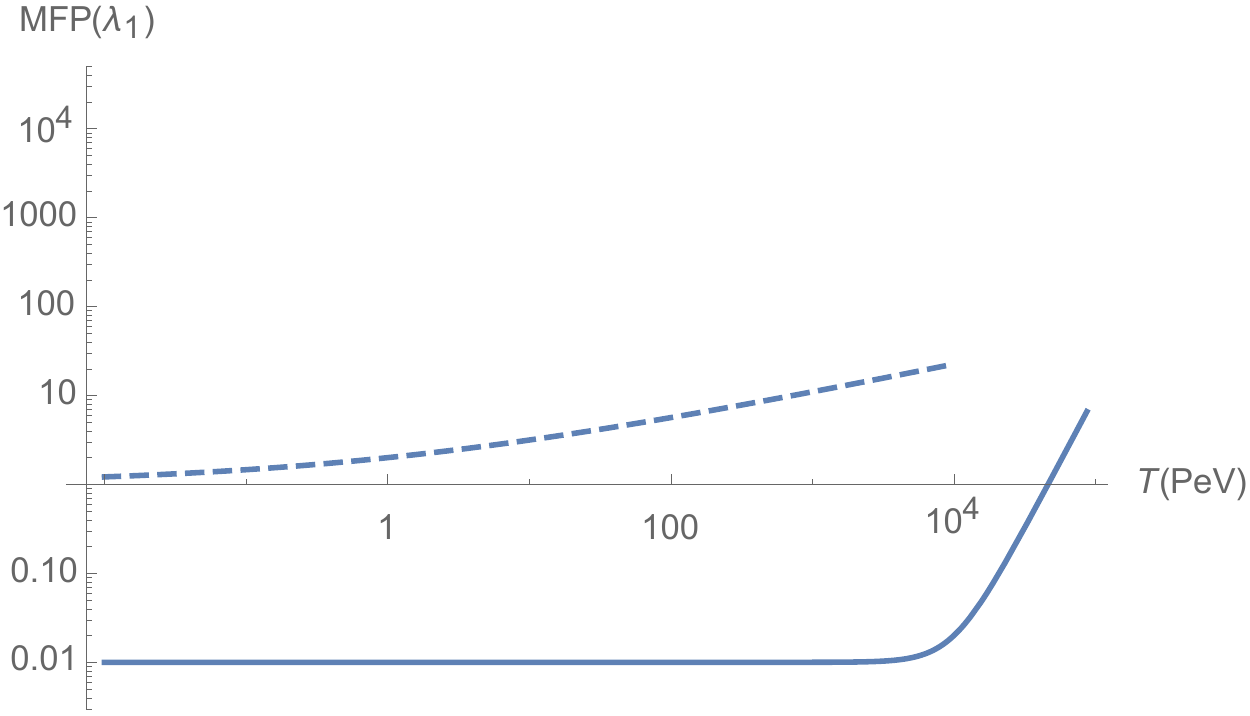}
\caption{The mean free path for protons versus the particle energy for undamped isotropic wave turbulence (dashed line) and damped case (solid line). Energy at which the mean free path change dependance on energy is $T_{c} \sim10^{19}eV$. For slab turbulence at $T_{H}\sim10^{15}eV$ the mean free path becomes infinitely large; in the isotropic turbulence that value is enhanced by factor $v/V_{A}$ due to resonant interaction - gyroresonace for undamped case and TTD resonant interaction in damped case. The second one provides even smaller mean free path due to the broadened resonance function. Mean free path is normalized by $\lambda_{1}$ in pc.}
\label{fig2}
\end{figure}

Moreover, as the cosmic rays mean free path in case of spatial gradients is closely related to the cosmic rays anisotropy \citep{Sch89}, the Hillas limit \citep{Hill84}
implies strong anisotropies at energies above 4 PeV which have not been yet confirmed by any experiment.

\section{Conclusions}
The implications of isotropically distributed
interstellar magnetohydrodynamic plasma waves together with damping effects on the scattering mean free path and consequently the spatial anisotropy of high-energy
cosmic rays has been investigated considering small pitch-angle scattering problem.
In the case of slab turbulence CR with Larmor radius $R_{L}$ can resonantly interact with certain plasma waves at $k_{res}\sim R_{L}^{-1}/\mu$ which is approximated for small pitch-angle as $k_{res}\sim R_{L}^{-1}$. We have already shown that this approximation is not valid. 
Drastic modification of the energy dependence of mean free path and consequently anisotropy
compared to previous calculations that have assumed that the plasma
waves propagate only parallel or antiparallel to the ordered magnetic field (slab turbulence) has been achieved. In the slab turbulence power spectrum vanishes for wavenumbers less than $k_{min}$,
since cosmic rays with Larmor radii larger
than $k^{-1}$ cannot be scattered in pitch-angle, which results in the so-called Hillas limit for the maximum energy of cosmic rays being confined in the Galaxy. For particles energies higher than this value spatial anisotropy becomes infinitely large that has not been detected by experiments considering UHECR. 

For isotropically distributed interstellar magnetohydrodynamic waves we demonstrated:

\begin{itemize}
  \item gyroresonance at resonant wave number  $k_{res} = (R_{L} \epsilon )^{-1}$ for $\mu=0$ is dominant in the case of undamped plasma waves;
  \item TTD resonant interaction is dominant for $\mu=0$ in the damped case due to resonance function broadening;
  \item particle energy limit that can be scattered and therefore confined is enhanced by four orders of magnitude $E_{c}\sim 10^{5}An^{1/2}(L /10 pc) PeV$ comparing to Hillas energy;
  \item mean free path is two orders of magnitude smaller comparing to undamped case.
\end{itemize} 

Below limiting energy the cosmic ray mean free path and the anisotropy exhibit the well known $E^{1/3}$ energy 
dependence, for $q = 5/3$ denoting the spectral index of the Kolmogorov spectrum for undamped plasma turbulence, while in the damped turbulence both transport parameters remain constant. 
At energies higher than $E_{c}$ mean free path and anisotropy steepen to a $E^{3}$-dependence. This implies that cosmic rays even close to ultrahigh energies of several tens of $EeV$ can be rapidly pitch-angle scattered by interstellar plasma turbulence, and are thus confined to the Galaxy.

\section*{Acknowledgements}
Part of this research is supported by the Ministry of Education and Science of the Republic of Serbia (contract 451-03-9/2021-14/200002).







\end{document}